\documentclass[10pt,twocolumn,showpacs,amsmath,amssymb,floatfix,superscriptaddress]{revtex4-2}

\usepackage{graphicx}
\usepackage{dcolumn}
\usepackage{bm}
\usepackage{color}
\usepackage{txfonts}
\usepackage{microtype}
\usepackage{hyperref}
\usepackage[english]{babel}
\usepackage{slashed}
\usepackage{gensymb}
\usepackage{epsfig}
\usepackage[normalem]{ulem}
\usepackage{amsmath}
\usepackage{array}
\usepackage{booktabs}
\allowdisplaybreaks[4]

\begin{document}
\renewcommand{\figurename}{FIG}	

\title{Angle-dependent pair production in the polarized two-photon Breit-Wheeler process}

\author{Qian Zhao}
\affiliation{Ministry of Education Key Laboratory for Nonequilibrium Synthesis and Modulation of Condensed Matter, Shaanxi Province Key Laboratory of Quantum Information and Quantum Optoelectronic Devices, School of Physics, Xi'an Jiaotong University, Xi'an 710049, China}
\author{Yan-Xi Wu}
\affiliation{Ministry of Education Key Laboratory for Nonequilibrium Synthesis and Modulation of Condensed Matter, Shaanxi Province Key Laboratory of Quantum Information and Quantum Optoelectronic Devices, School of Physics, Xi'an Jiaotong University, Xi'an 710049, China}
\affiliation{Department of Physics, Northwest University, Xi'an 710069, China}
\author{Mamutjan Ababekri}
\affiliation{Ministry of Education Key Laboratory for Nonequilibrium Synthesis and Modulation of Condensed Matter, Shaanxi Province Key Laboratory of Quantum Information and Quantum Optoelectronic Devices, School of Physics, Xi'an Jiaotong University, Xi'an 710049, China}
\author{Zhong-Peng Li}
\affiliation{Ministry of Education Key Laboratory for Nonequilibrium Synthesis and Modulation of Condensed Matter, Shaanxi Province Key Laboratory of Quantum Information and Quantum Optoelectronic Devices, School of Physics, Xi'an Jiaotong University, Xi'an 710049, China}
\author{Liang Tang}\email{tangl@hebtu.edu.cn}
\affiliation{College of Physics and Hebei Key Laboratory of Photophysics Research and Application, Hebei Normal University, Shijiazhuang 050024, China}
\author{Jian-Xing Li}\email{jianxing@xjtu.edu.cn}
\affiliation{Ministry of Education Key Laboratory for Nonequilibrium Synthesis and Modulation of Condensed Matter, Shaanxi Province Key Laboratory of Quantum Information and Quantum Optoelectronic Devices, School of Physics, Xi'an Jiaotong University, Xi'an 710049, China}

\date{\today}
	
\begin{abstract}
The advent of laser-driven high-intensity $\gamma$-photon beams has opened up new opportunities for designing advanced photon-photon colliders. Such colliders have the potential to produce a large yield of linear Breit-Wheeler (LBW) pairs in a single shot, which offers a unique platform for studying the polarized LBW process. In our recent work [Phys. Rev. D 105, L071902(2022)], we investigated the polarization characteristics of LBW pair production in CP $\gamma$-photon collisions. To fully clarify the polarization effects involving both CP and LP $\gamma$-photons, here we further investigate the LBW process using the polarized cross section with explicit azimuthal-angle dependence due to the base rotation of photon polarization vectors. We accomplished this by defining a new spin basis for positrons and electrons, which enables us to decouple the transverse and longitudinal spin components of $e^\pm$. By means of analytical calculations and Monte Carlo simulations, we find that the linear polarization of photon can induce the highly angle-dependent pair yield and polarization distributions. The comprehensive knowledge of the polarized LBW process will also open up avenues for investigating the higher-order photon-photon scattering, the laser-driven quantum electrodynamic plasmas and the high-energy astrophysics.

\end{abstract}

	\maketitle
\section{Introduction}\label{introduction}
The quantum electrodynamics (QED) theory predicts the interaction between photons, wherein the collision of two real $\gamma$ photons can produce the $e^+e^-$ pair, known as the linear Breit-Wheeler (LBW) process. The LBW process is the second-order process in QED \cite{Breit1934}, and its validation in terrestrial laboratories requires an ultrahigh-brilliance $\gamma$-ray beam with a peak cross-section of $\sim1.6\times10^{-29}$ m$^2$ \cite{Zhao2022}. To date, the only verified interaction of real photons in experiments is multiphoton Breit-Wheeler pair production \cite{Burke1997}. However, the production of LBW pairs via virtual photons has been demonstrated in high-energy collider experiments by utilizing the photons of a highly Lorentz-contracted Coulomb field \cite{Adam2021}.

The LBW process is a fundamental ingredient in high-energy astrophysics, playing a crucial role in the production of pair plasma in $\gamma$-ray bursts \cite{Kumar2015,Kostenko2018,Lundman2018} and black hole activity \cite{Hirotani2016,Akiyama2019}. Furthermore, the LBW process can dominate the laser-driven QED plasmas, as demonstrated by numerical simulations \cite{He2021phys,He2021njp}. In recent years, several $\gamma\gamma$ colliders have been proposed in the platform of laser-plasma interactions to produce LBW pairs \cite{Pike2014,Ribeyre2016,Drebot2017,Jansen2018,Yu2019,Wang2020,Golub2021,Kettle2021,Esnault2021}.

Despite the calculation of the total LBW cross section $\sigma_{\gamma\gamma}$ for increasing the luminosity of $\gamma\gamma$ colliders, the photon polarization and $e^\pm$ spins, which are the fundamental quantum nature of the photons and leptons \cite{Sun2022}, entail the exclusive characteristics of the LBW process \cite{Budnev1975,Ginzburg1984,Ivanov2005,Bakmaev2007,Harland-lang2019,Adam2021,Zhao2022}. When utilizing two linearly polarized (LP) photons, $\sigma_{\gamma\gamma}$ can be expressed in terms of mutual-parallel or mutual-perpendicular polarization vectors, and the non-polarized cross section can be obtained via a simple average over the two LP cross sections \cite{Budnev1975,Harland-lang2019}. The LBW process has been demonstrated in high-energy colliders by means of the equivalent photon approximation of ultra-relativistic heavy-ions, which presents the first measurement of the unique $\cos{4\varphi}$ modulation of the distinct differential cross section with two LP photons, where $\varphi$ is the angle between the lepton pair transverse momentum and the individual lepton transverse momentum \cite{Li2019probing,Li2020impact}. Fundamentally, the angular modulations, both in polar and azimuthal directions, originate from the total helicity of the $\Lambda_\gamma=0,\pm2$ two-photon system, which must be transferred to the orbital angular momentum of the $e^+e^-$ pair \cite{Adam2021,Zhao2022}. Therefore, the LBW differential cross section can be extended to the fundamental level of helicity amplitudes to more deeply investigate the effects of photon polarization \cite{Brandenburg2022}.

Moreover, to clarify the effects of arbitrary photon polarization, the polarization states of photons and pairs can be described by the corresponding density matrices, leading to the completely polarized LBW cross section in terms of Stokes parameters, which is applicable to perform the Monte Carlo (MC) simulation incorporating the beam effects \cite{Zhao2022,Lu2022}. The collision of quasi-energetic circularly polarized (CL) $\gamma$-photon beams can produce distinct energy-angle spectra. When the colliding photons have the same right-hand or left-hand helicities, the collision produces the quadrupole angular spectrum imprinted by the longitudinal polarization. Conversely, the opposite helicities lead to the dipole one imprinted by the transverse polarization \cite{Zhao2022}. Although the effects of independent linear polarization or circular polarization have been elucidated, their coupling effects are still elusive when the colliding photons are partially circular polarization and linear polarization.

In this paper, we undertake an investigation of the fully angle-dependence in the polarized LBW process using both analytical cross-section and MC numerical simulation. By defining a new spin base, the transverse and longitudinal polarization of $e^\pm$ are decoupled from the differential cross section, allowing for the analytical clarification of the effects of photon linear polarization on the angle-distribution of $e^+e^-$ pair polarization. By utilizing the MC simulation, we have quantitatively retrieved the azimuthal and polar dependence of $e^+e^-$ pair, revealing that the linear polarization of photons induces an azimuthal-dependent pair yield and pair polarization. This comprehensive understanding of the polarized LBW process is highly beneficial for the upcoming $\gamma\gamma$ collider and the associated research in high-energy astrophysics.

The paper is structured as follows. In Sec.~\ref{method}, the theoretical deduction of the completely polarized LBW cross-section is described, and the method of spin-resolved MC sampling is established. The effects of arbitrary photon polarization are analyzed theoretically and numerically in Sec.~\ref{results}. The proposed  ultrabrilliant polarized $\gamma$-photon source that could be used to perform the polarized LBW verification is  discussed in \ref{discussion}.
Finally, a brief conclusion of this paper is presented in Sec.~\ref{summary}.

\section{Methods of theory and numerical simulation}\label{method}
The LBW process's polarized scattering amplitude, with arbitrary initial photon polarization and final positron (electron) spins, can generally be described through the photon and lepton density matrixes. Stokes parameters and mean spin vectors can be utilized to describe the arbitrary polarization states of the photon and $e^\pm$ in the cross section. To establish a connection between theoretical predictions and experimental procedures, a numerical simulation method was developed using MC sampling, which permits the consideration of beam effects stemming from the energy spread and initial polarization distributions of photon beams. The no-time-count method was employed to implement the numerical simulation, converting the macro beam-beam collision into the micro single scattering \cite{Gaudio2020}. Our MC simulation strategy for binary collision is described in detail in \cite{Zhao2022,Lu2022}.
\subsection{Calculation of completely spin-dependent LBW cross section}\label{theory}

In the calculation of the completely spin-dependent cross section of the LBW process, we use the Lorentz invariant density matrix to describe the arbitrary polarization.
The Mandelstam invariants in the LBW process are defined as
\begin{subequations}\label{stu}
\begin{eqnarray}
s&=&(k_1+k_2)^2/m_e^2=(p_++p_-)^2/m_e^2,\\
t&=&(k_1-p_+)^2/m_e^2=(p_--k_2)^2/m_e^2,\\
u&=&(k_1-p_-)^2/m_e^2=(p_+-k_2)^2/m_e^2,
\end{eqnarray}
\end{subequations}
where $m_e$ is the mass of electron, $k_{1,2}$ the 4-momenta of $\gamma^{(1,2)}$ and $p_\pm$ the 4-momenta of $e^\pm$. The LBW cross section is formulated in the center of mass (c.m.) frame as (relativistic units with $\hbar=c=1$ are used throughout)
\begin{eqnarray}
\frac{d\sigma_{\gamma\gamma}}{d\Omega}=\frac{r_e^2m_e^2|\bm{p}_e|}{16\varepsilon^3}|\epsilon_1^\mu M_{\mu\nu}\epsilon_2^\nu|^2,\label{dsigBW}
\end{eqnarray}
where $\Omega$ is the solid angle, $r_e$ the classical electron radius,  $\varepsilon$ and $\bm{p}_e$ the c.m. energy and c.m. momentum of electron, respectively. Defining  $M_{fi}=\epsilon_1^\mu \epsilon_2^\nu M_{\mu\nu}$, the scattering amplitude $|M_{fi}|^2$ is expressed as
\begin{eqnarray}\label{scatamp}
|M_{fi}|^2&=&\mid\bar{u}(p_-,s_-)\epsilon_1^\mu(k_1,\lambda_1) \epsilon_2^\nu(k_2,\lambda_2) Q_{\mu\nu}v(p_+,s_+)\mid^2, \nonumber\\
&=&[\bar{u}\epsilon_1^{\mu}\epsilon_1^{*\rho}Q_{\mu\nu} v][\bar{v}\epsilon_2^{\nu}\epsilon_2^{*\sigma}\bar{Q}_{\rho\sigma} u],\nonumber\\
&=&{\rm{Tr}^{\rm{Dirac}}}[u\bar{u}(\epsilon_1^{\mu}\epsilon_1^{*\rho}Q_{\mu\nu})v\bar{v}(\epsilon_2^{\nu}\epsilon_2^{*\sigma}\bar{Q}_{\rho\sigma})],\label{traceMfi}
\end{eqnarray}
where $s_\pm$ are the spin 4-vectors of $e^\pm$, $\lambda_{1,2}$ the photon helicities, and $Q_{\mu\nu}$ the Lorentz tensor
\begin{equation}
Q_{\mu\nu}=\gamma_\mu\frac{-\slashed{p}_++\slashed{k}_2+m}{t-m^2}\gamma_\nu+\gamma_\nu\frac{-\slashed{p}_++\slashed{k}_1+m}{u-m^2}\gamma_\mu.
\end{equation}
The scattering amplitude at mixed state is calculated by changing the tensor to density matrix, i.e., $u\bar{u}\rightarrow\rho^-, v\bar{v}\rightarrow\rho^+, \epsilon_1^\mu\epsilon_1^{*\rho}\rightarrow\rho^{(1)\mu\rho}$, and $\epsilon_2^\nu\epsilon_2^{*\sigma}\rightarrow\rho^{(2)\nu\sigma}$.

By using of the orthogonal 4-momenta (see \cite{Berestetskii1982},\S87):
\begin{subequations}\label{orthovec}
\begin{eqnarray}
Q&=&k_1+k_2=p_-+p_+,\\
K&=&k_1-k_2,\\
P_\perp&=&P-\frac{KP}{K^2}K,\\
N^\lambda&=&\upsilon^{\lambda\mu\nu\rho}Q_{\mu}K_{\nu}P_{\rho},
\end{eqnarray}
with $P=p_+-p_-$ and Levi-Civita tensor $\upsilon^{\lambda\mu\nu\rho}$.
\end{subequations}
It is convenient to define an orthogonal basis
\begin{eqnarray}\label{photbasis}
e_1=-N/\sqrt{-N^2}, ~~~e_2=P_\perp/\sqrt{-P_\perp^2},
\end{eqnarray}
by which the density matrix of photon can be expanded by the Stokes parameter $\bm{\xi}$ as
\begin{equation}\label{photden}
\rho^{(\gamma)\mu\nu}=\frac{1}{2}(1+\bm{\xi}\bm{\sigma})_{ll'}e^\mu_{l}e^\nu_{l'},~~~l,l'=1,2.
\end{equation}
The spin basis can be constructed from the orthogonal vectors in Eq. (\ref{orthovec})
\begin{eqnarray}\label{ebase}
\left\{
\begin{array}{lr}
  e^{\pm}_0=p_\pm/{m_{e}}~,& \\
  e^\pm_1=N/{\sqrt{-|N|^2}}~,& \\
  e^{\pm}_2=(f_1p_\pm+f_2p_\mp)/m_e~,&\\
  e^{\pm}_3=(f_3K+f_4P_\perp)/m_e~,&
\end{array}
\right.
\end{eqnarray}
where the coefficients $f_1,f_2,f_3$ and $f_4$ are determined by the orthogonal relations between $e^{\pm}_2$ and $e^{\pm}_3$, and the normalization condition $|e^{\pm}_2|^2=|e^{\pm}_3|^2=-1$. With the spin basis it has $a_\pm=\sum_{i=1}^{3}\zeta_i^\pm e_i^\pm$ with mean spin 3-vector $\bm{\zeta}^\pm$, and the density matrixes,  $\rho_\pm=1/(2m_e)(\slashed{p}_\pm\mp m_e)[1-\gamma^5(\slashed{s}_\pm)]$, can be expanded as
\begin{equation}\label{denmatrix}
\rho^\pm=\frac{1}{2}\sum_{i=0}^{3}\zeta_i^\pm\rho_i^\pm,
\end{equation}
where $\zeta_0=1$, $\rho_0^\pm=\slashed{p}_\pm\mp m$ and $\rho_i^\pm=-\rho_0^\pm\gamma_5\slashed{e}_i^\pm$. Substituting Eqs. (\ref{photden}) and (\ref{denmatrix}) into Eq. (\ref{scatamp}) and performing the calculation of Dirac trace \cite{Shtabovenko2020}, one obtains the differential cross section of the LBW process
\begin{equation}\label{difcs}
\frac{d^2\sigma_{\gamma\gamma}}{dtd\varphi}=\frac{r_e^2m_e^2}{64\varepsilon^4}\left[F+\sum_{i=1}^{3}(G_i^+\zeta_i^++G_i^-\zeta_i^-)+\sum_{i,j=1}^{3}H_{i,j}\zeta_i^+\zeta_j^-\right],
\end{equation}
where the relation $d\Omega$=-$dt$$d\varphi$/2$|\bm{p}_e||\bm{k}|$ is used, and the functions $F$, $G_i^\pm$ and $H_{i,j}$ are expressed by the normalized Mandelstam invariants and photon Stokes parameters (see Appendix \ref{appB}). With the spin summation of final $e^+$ and $e^-$ in Eq. (\ref{difcs}), one obtains the spin-summarized differential cross section $d\bar{\sigma}_{\gamma\gamma}/dtd\varphi$ [see Eq. (\ref{dsig_spinsum})].

Polarization vector of final $e^\pm$ in the scattering process itself is (see \S65 in \cite{Berestetskii1982})
\begin{eqnarray}\label{zeta}
\zeta^{(f)}_{+,i}=\frac{G^+_{i}}{F},~~~ \zeta^{(f)}_{-,i}=\frac{G^-_{i}}{F}, ~~~i=1,2,3.
\end{eqnarray}
In the c.m. frame, $\zeta^\pm_1$ is the transverse polarization perpendicular to the scattering plane, $\zeta^\pm_3$ is the transverse polarization in the scattering plane, and $\zeta^\pm_2$ is the longitudinal polarization (see details in Sec. \ref{simulation}).  Denoting $\gamma_{R}^{(1,2)}$ as right-hand circular polarization and $\gamma_{L}^{(1,2)}$ as left-hand circular polarization photon, For $\gamma^{(1)}_R\gamma^{(2)}_L$ collision, the corresponding helicity amplitudes can be obtained by setting $\zeta^\pm_2=\pm1$ in Eq.~(\ref{difcs})
\begin{subequations}\label{4helicity}
\begin{eqnarray}
|M_{+-\pm\mp}|^2&=&F\pm G^-_{2}\mp G^+_{2}-H_{22}~,\\
|M_{+-\pm\pm}|^2&=&F\pm G^-_{2}\pm G^+_{2}+H_{22}~.
\end{eqnarray}
\end{subequations}
The differential cross sections with the four helicity channels $|+-\pm\mp\rangle$ and $|+-\pm\pm\rangle$ are denoted as $d\sigma_{+-\pm\mp}$ and $d\sigma_{+-\pm\pm}$, respectively. In the subscripts, the first to the fourth position correspond to the positive (``$|+\rangle$'') or negative (``$|-\rangle$'') helicity eigenstates of $\gamma^{(1)}, \gamma^{(2)}, e^+$ and $e^-$, respectively. Additionally, the helicity scattering amplitudes can be obtained directly by utilizing the helicity states in the calculation of $\mid M_{fi}\mid^2$ with spin 4-vectors $s_\pm=\lambda_\pm\left(\frac{|\bm{p}|}{m_e},\frac{E_\pm}{m_e}\frac{\bm{p}}{|\bm{p}|}\right)$. This procedure is consistent with the deduced ones from the arbitrarily-polarized scattering amplitude in Eq. (\ref{difcs}).

The spin 3-vectors ${\bm{\zeta}}^{(f)}_{\pm}$ can be expressed in an arbitrary frame by the definition of  a set of 3-vector basis $\bm{n}_{\pm}$ \cite{Kotkin1998}
\begin{eqnarray}\label{spinvec}
\bm{\zeta}^{(f)}_{\pm}&=&\sum_{i=1}^{3} \zeta^{(f)}_{\pm,i} \bm{n}_{\pm,i},\\
\bm{n}_{\pm,i}&=&\bm{e}_{i}^{\pm}- \bm{p}_\pm/(E_\pm+m_e) e_{i0}^{\pm},
\end{eqnarray}
with  $e_{i0}^{\pm}$ being a time-component of 4-vector $e_{i}^{\pm}$ defined in Eq. (\ref{ebase}), and, $E_\pm$ and $\bm{p}_\pm$ being the energies and momenta of $e^{\pm}$ in an arbitrary frame. Thus, the mean helicities of $e^+e^-$ pair are expressed as $\lambda_\pm=\bm{\zeta}^{(f)}_{\pm}\bm{p}_\pm/(2| \bm{p}_\pm|)$ in an arbitrary frame.

\subsection{Constructing polarized LBW process with MC sampling}\label{simulation}
\begin{figure}[!t]
	\setlength{\abovecaptionskip}{0.2cm}
	\centering\includegraphics[width=0.8\linewidth]{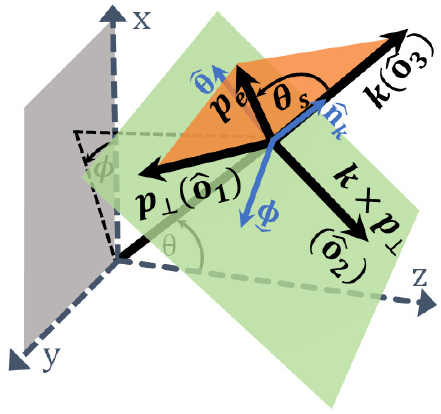}
	\caption{Coordinate system of the LBW process in the c.m. frame. $\bm{p}_e$ and $\bm{k}$ are the momenta of electron and photon $\gamma^{(1)}$. $(\hat{n},\hat{\theta},\hat{\phi})$ compose the spherical coordinates, and $\hat{\theta}$ and $\hat{\phi}$ are polar angle and azimuth angles of $\bm{k}$. ($\hat{\bm{o}}_1,\hat{\bm{o}}_2,\hat{\bm{o}}_3$) compose the spherical coordinates. $\bm{p}_{\perp}$ is perpendicular to $\bm{k}$ and onto the plane of $\hat{\bm{o}}_1$ and $\hat{\bm{o}}_2$. $\bm{p}_e$ is the momentum of electron and onto the plane of $\hat{\bm{o}}_1$ and $\hat{\bm{o}}_3$.}
	\label{fig:mom}
\end{figure}

Firstly, a clarification of the coordinate system of the LBW process in the c.m. frame is presented in Fig. \ref{fig:mom}. In this frame, the polar and azimuthal directions of a pair are defined with respect to the photon momentum $\bm{k}$. Specifically, the polar direction is given by the scattering angle $\theta_s$ in the scattering plane, while the azimuthal angle $\varphi$ is determined by $\bm{p}_\perp$ in the azimuthal plane $\hat{\bm{\theta}}-\hat{\bm{\phi}}$. The scattering process is thus described in the coordinate system ($\hat{\bm{o}}_1,\hat{\bm{o}}_2,\hat{\bm{o}}_3$).

Upon Lorentz boost along the c.m. frame velocity $\bm{\beta}_{cm}=(\bm{k}_1+\bm{k}_2)/(\omega_1+\omega_2)$, the colliding photons are transformed into the c.m. frame. Specifically, $\gamma^{(1)}$ photon has 4-momentum $(\varepsilon,\bm{k})$, while $\gamma^{(2)}$ photon has 4-momentum $(\varepsilon,-\bm{k})$ with $\varepsilon=\sqrt{\omega_1\omega_2(1-\cos{\theta_i})/2}$. Here, $\omega_1$ and $\omega_2$ ($\bm{k}_1$ and $\bm{k}_2$) denote the energies (wave vectors) of $\gamma^{(1)}$ and $\gamma^{(2)}$ photons in the laboratory frame, respectively, and $\theta_i$ is the angle between $\bm{k}_1$ and $\bm{k}_2$.

In the context of the LBW process, the initial polarization vector of a photon can be specified in the basis of its momentum. In the c.m. frame, the polarization basis of $\gamma^{(2)}$ must first be transformed to that of $\gamma^{(1)}$, which involves modifying the Stokes parameters of $\gamma^{(2)}$ as follows: $\xi_1^{(2)}\rightarrow-\xi_1^{(2)},\xi_2^{(2)}\rightarrow-\xi_2^{(2)},\xi_3^{(2)}\rightarrow\xi_3^{(2)}$. Subsequently, the 4-vector basis in Eq. (\ref{photbasis}) only retains the vector components and corresponds to $\hat{\bm{o}}_1$ and $\hat{\bm{o}}_2$, as depicted in Fig. \ref{fig:mom}, which are given by
\begin{align}
 \hat{\bm{o}}_1=\frac{\bm{p}_\perp}{|\bm{p}_\perp|},~~ \hat{\bm{o}}_1=\frac{\bm{k}\times\bm{p}_\perp}{|\bm{k}\times\bm{p}_\perp|}.
 \end{align}
To take into account the effects of linear polarization of photons, the Stokes parameters $\bm{\xi}^{(1)}$ and $\bm{\xi}^{(2)}$, which are defined in the basis ($\hat{\theta}, \hat{\phi}$), must be transformed to the basis $(\bm{o}^{(1)},\bm{o}^{(2)})$ with a counter-clockwise rotation by an angle $\varphi$. The resulting transformed Stokes parameters are given by Eqs. (\ref{eta1}) and (\ref{eta2}), which read
\begin{subequations}\label{eta1}
\begin{align}
 \eta_1^{(1)}=&\xi _1^{(1)}\cos{2\varphi}-\xi _3^{(1)}\sin{2\varphi},\\
 \eta_3^{(1)}=&\xi _1^{(1)}\sin{2\varphi}+\xi _3^{(1)}\cos{2\varphi}, \\
 \eta_2^{(1)}=&\xi _2^{(1)},
\end{align}
\end{subequations}
and
\begin{subequations}\label{eta2}
\begin{align}
 \eta_1^{(2)}=&-\xi _1^{(2)}\cos{2\varphi}-\xi _3^{(2)}\sin{2\varphi},\\
 \eta_3^{(2)}=&-\xi _1^{(2)}\sin{2\varphi}+\xi _3^{(2)}\cos{2\varphi},\\
 \eta_2^{(2)}=&-\xi _2^{(2)},
\end{align}
\end{subequations}
where $\varphi$ represents the azimuthal angle of the electron in the spherical basis ($\hat{\bm{n}}_k,\hat{\bm{\theta}},\hat{\bm{\phi}})$. The transformed Stokes parameters in Eqs. (\ref{eta1}) and (\ref{eta2}) can then be substituted into Eq. (\ref{difcs}) to account for the linear polarization effects of photons.

In the LBW process, the c.m. momenta of the $e^\pm$ particles are determined within the defined coordinate system of Fig. \ref{fig:mom} via the relations given in Eqs. (\ref{npe}), which can be expressed as
\begin{subequations}\label{npe}
\begin{eqnarray}
 \hat{\bm{n}}_\perp&=&\cos\varphi\hat{\bm{\theta}}+\sin\varphi\hat{\bm{\phi}},\\
 \hat{\bm{n}}_e&=&\cos{\theta_s}\hat{\bm{n}}_k+\sin{\theta_s}\hat{\bm{n}}_\perp.
\end{eqnarray}
\end{subequations}
Here, $d\bar{\sigma}_1$ and $d\bar{\sigma}_2$ are defined as integrals of $d\bar{\sigma}_{\gamma\gamma}/dtd\varphi$ with respect to $\varphi$ and $t$, respectively, and are used to determine the scattering angle $\theta_s$ and the azimuthal angle $\varphi$. Specifically, the values of $\theta_s$ and $\varphi$ are obtained by solving Eqs. (\ref{thephi}), which can be written as
\begin{subequations}\label{thephi}
\begin{align}
\int_{-|\cos{\theta_s}|}^{|\cos{\theta_s}|}d\bar{\sigma}_1=\frac{\sigma_{\gamma\gamma}}{\sqrt{s(s-4)/4}}|R_1|,~~
\int_{0}^{\varphi}d\bar{\sigma}_2=\sigma_{\gamma\gamma}|R_2|,
\end{align}
\end{subequations}
where $R_1$ and $R_2$ are uniform random numbers between -1 and 1, and 0 and 1, respectively. The total cross section $\sigma_{\gamma\gamma}$ is obtained by integrating $d\bar{\sigma}_{\gamma\gamma}/dtd\varphi$ over $\varphi$ (between $0<\varphi<2\pi$) and $t$ (between $m_e^2-2\omega^2\pm2\omega\sqrt{\omega^2-m_e^2}$) [see Eq. \ref{sig_tot}].

In the c.m. frame, the basis vectors of the spin 3-vector defined in Eq.~(\ref{spinvec}) correspond to the directions of the spherical coordinates $(\hat{\bm{\phi}}_{p},\hat{\bm{n}}_{p},\hat{\bm{\theta}}_{p})$ of $\bm{p}_e$, as shown in Fig.\ref{fig:mom}. As a result, in the c.m. frame, each single reaction produces an $e^-e^+$ pair with longitudinal polarization $\zeta^{(f)}_{\pm,2}$ and transverse polarization $\zeta^{(f)}_{\pm,1}$ and $\zeta^{(f)}_{\pm,3}$ along $\hat{\bm{\phi}}_{p}$ and $\hat{\bm{\theta}}_{p}$, respectively. We denote $\zeta_{\parallel}$ and $\zeta_{\perp}$ as the longitudinal and transverse polarizations, respectively, which can be expressed as
\begin{equation}\label{poldef}
 \zeta_{\parallel}\equiv\zeta^{(f)}_{\pm,2},~~~ \zeta_{\perp}\equiv\sqrt{\left(\zeta^{(f)}_{\pm,1}\right)^2+\left(\zeta^{(f)}_{\pm,3}\right)^2}. \end{equation}
Thus, the energy and angle-dependent polarization can be obtained analytically using Eq. (\ref{poldef}). Alternatively, $\zeta_{\parallel}$ can be calculated as the difference between the counting rates for positive and negative helicities (summed over polarization states of $e^+$), normalized to the total counting rate, as given by
\begin{eqnarray}\label{polsig}
\zeta_{\parallel}=\frac{\sum\limits_{\bm{\zeta}_{+}} \left[d\sigma_{\gamma\gamma}(\zeta_{-,2}=1)-d\sigma_{\gamma\gamma}(\zeta_{-,2}=-1)\right]}{\sum\limits_{\bm{\zeta}_{+}}\left[d\sigma_{\gamma\gamma}(\zeta_{-,2}=1)+d\sigma_{\gamma\gamma}(\zeta_{-,2}=-1)\right]},
\end{eqnarray}
and a similar expression can be derived for $\zeta_\perp$ based on the counting rates.

The determination of the single-spin state can be achieved using MC sampling based on the transition probabilities given in Eq. (\ref{difcs}). The statistical interpretation of the mean polarization in Eq. (\ref{poldef}) is that the distribution of the mean polarization in the phase space can be obtained from the averaged spin components observed by a detector. The observed spin states of the $e^+e^-$ pair are determined by four transition probabilities
\begin{subequations}\label{spinpro}
	\begin{align}
	W^{\uparrow\uparrow}&=\int d\Omega\left[1/4+\bm{\zeta}^{(f)}_+\bm{\zeta}_+^{(d)}+\bm{\zeta}^{(f)}_-\bm{\zeta}_-^{(d)}+(\bm{\zeta}_-^{(d)})^TH\bm{\zeta}_+^{(d)}\right],\\
	W^{\uparrow\downarrow}&=\int d\Omega\left[1/4+\bm{\zeta}^{(f)}_+\bm{\zeta}_+^{(d)}-\bm{\zeta}^{(f)}_-\bm{\zeta}_-^{(d)}-(\bm{\zeta}_-^{(d)})^TH\bm{\zeta}_+^{(d)}\right],\\
	W^{\downarrow\uparrow}&=\int d\Omega\left[1/4-\bm{\zeta}^{(f)}_+\bm{\zeta}_+^{(d)}+\bm{\zeta}^{(f)}_-\bm{\zeta}_-^{(d)}-(\bm{\zeta}_-^{(d)})^TH\bm{\zeta}_+^{(d)}\right],\\
	W^{\downarrow\downarrow}&=\int d\Omega\left[1/4-\bm{\zeta}^{(f)}_+\bm{\zeta}_+^{(d)}-\bm{\zeta}^{(f)}_-\bm{\zeta}_-^{(d)}+(\bm{\zeta}_-^{(d)})^TH\bm{\zeta}_+^{(d)}\right].
	\end{align}
\end{subequations}
which consist of the spin-projecting terms (the second and third terms in square bracket) and the correlation term (the fourth term in square bracket) between $\bm{\zeta}^{(f)}$ and the spin axis $\bm{\zeta}^{(d)}$ (a unit vector) of a detector. The observed spin components $\zeta_{\pm,i}^{(d)}=\zeta^{(f)}{\pm,i}/|\bm{\zeta}^{(f)}{\pm}|$ ($i=1,2,3$) are determined by each transition probability, with either a parallel or anti-parallel projection onto the spin axis. MC sampling, using a uniform random number $0<R_3<1$, can be employed to determine the observed spin components. The values of $R_3$ determine the observed spin components as follows: $+\zeta_{+,i}^{(d)},+\zeta_{-,i}^{(d)}$ if $R_3\in(0,W^{\uparrow\uparrow})$; $-\zeta_{+,i}^{(d)},-\zeta_{-,i}^{(d)}$ if $R_3\in(W^{\uparrow\uparrow},W^{\downarrow\downarrow}+W^{\uparrow\uparrow})$; $+\zeta_{+,i}^{(d)},-\zeta_{-,i}^{(d)}$ if $R_3\in(W^{\downarrow\downarrow}+W^{\uparrow\uparrow},W^{\uparrow\downarrow}+W^{\downarrow\downarrow}+W^{\uparrow\uparrow})$; and $-\zeta_{+,i}^{(d)},+\zeta_{-,i}^{(d)}$ if $R_3\in(W^{\uparrow\downarrow}+W^{\downarrow\downarrow}+W^{\uparrow\uparrow},1)$.
Defining $\bm{\zeta}_\pm^{(d)}$ along the directions of $\bm{\zeta}^{(f)}_\pm$ leads to the statistical total polarization of the produced $e^\pm$ beams
\begin{align}\label{polbeam}
P_{tot}=\sqrt{(\bar{\zeta}_{\pm,1}^{(d)})^2+(\bar{\zeta}_{\pm,2}^{(d)})^2+(\bar{\zeta}_{\pm,3}^{(d)})^2},
\end{align}
where $\bar{\zeta}_{\pm,i}^{(d)}$ are the averaged components over the particle number. If $\bm{\zeta}\pm^{(d)}$ is defined as the parallel or perpendicular directions of the $e^\pm$ momenta, one obtains the statistical longitudinal ($P_{\parallel}$) or transverse polarization ($P_{\perp}$).

\section{Theoretical and numerical analysis of polarization effects in LBW process}\label{results}
\subsection{Analytical calculation of pair polarization from angle-dependent cross section}
\begin{figure}[!t]
	\setlength{\abovecaptionskip}{0.2cm}
	\centering\includegraphics[width=1\linewidth]{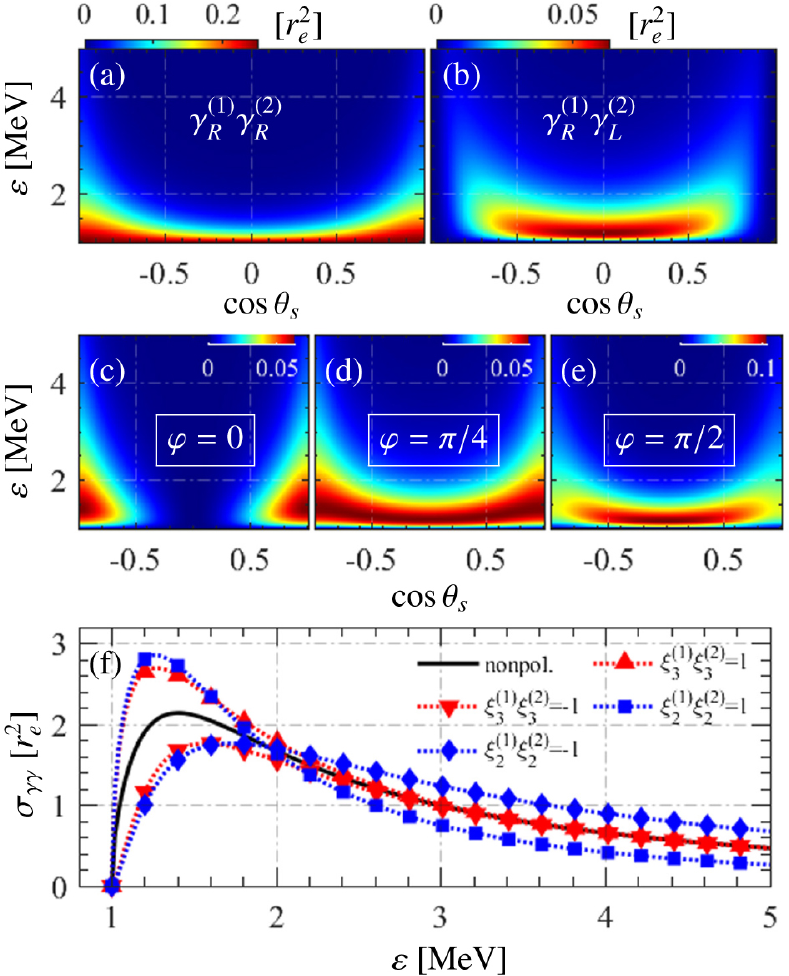}
	\caption{(a) and (b) Distributions of $d\bar{\sigma}_{\gamma\gamma}/dtd\varphi$ with $\gamma_R^{(1)}\gamma_R^{(2)}$ and $\gamma_R^{(1)}\gamma_L^{(2)}$ collisions, respectively. (c)-(e) Distributions of $d\bar{\sigma}_{\gamma\gamma}/dtd\varphi$ with $\xi^{(1)}_3=\xi^{(2)}_3=1$ corresponding to different azimuthal angles $\varphi$. (f) The total cross section $\sigma_{\gamma\gamma}$ of the LBW process with nonpolarized (nonpol.) photons and different photon polarizations.}
	\label{fig:dsig}
\end{figure}
\begin{figure}[!t]
	\setlength{\abovecaptionskip}{0.2cm}
	\centering\includegraphics[width=1\linewidth]{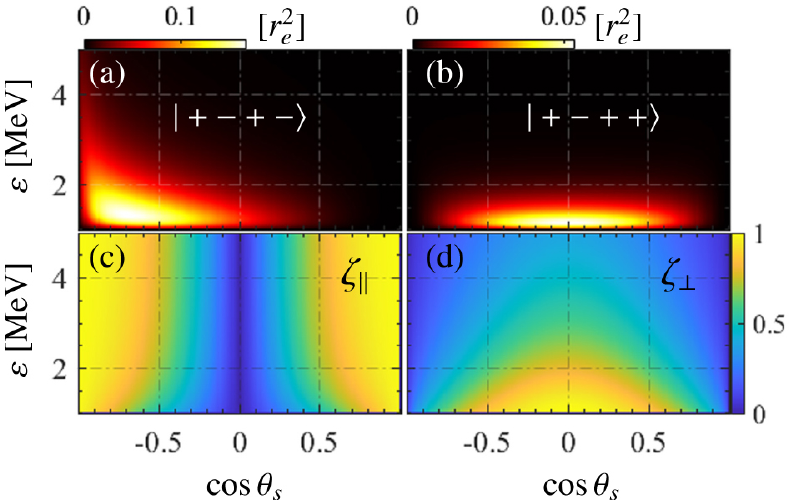}
\caption{In the scenario of $\gamma_R^{(1)}\gamma_L^{(2)}$ collision:  (a) and (b) Distributions of $d\sigma_{\gamma\gamma}/dtd\varphi$ with $|+-+-\rangle$ and $|+-++\rangle$ helicity channels.  (c)-(d) Distributions of longitudinal $\zeta_\parallel$ and transverse $\zeta_\perp$ polarization, calculated from Eq. (\ref{poldef}).}
	\label{fig:helicity}
\end{figure}

The dependence of the spin-summarized cross section $d\bar{\sigma}_{\gamma\gamma}/dtd\varphi$ on energy and angle is resolved by different configurations of photon polarization, as illustrated in Figure \ref{fig:dsig}. The expression of $d\bar{\sigma}_{\gamma\gamma}/dtd\varphi$, given by Eq. (\ref{dsig_spinsum}), reveals that $d\bar{\sigma}_{\gamma\gamma}/dtd\varphi$ is independent of $\varphi$ for the CP photons ($\xi_2^{(1)}\xi_2^{(2)}=\pm1$). The distribution of $d\bar{\sigma}_{++}/dtd\varphi$ indicates that dominant reactions occur along the colliding axis ($\cos{\theta_s}=\pm1$) in the $\gamma_R^{(1)}\gamma_R^{(2)}$ collision, when the energy is far beyond the threshold [see Figure \ref{fig:dsig} (a)]. In contrast, reactions are almost forbidden near the colliding axis in the $\gamma_R^{(1)}\gamma_L^{(2)}$ collision [see Figure \ref{fig:dsig} (b)]. The distinct energy-angle dependence in $\gamma_R^{(1)}\gamma_R^{(2)}$ and $\gamma_R^{(1)}\gamma_L^{(2)}$ collisions can be attributed to the angular momentum conservation of the helicity transfer [see Figure \ref{fig:helicity}]. For the LP photons, $d\bar{\sigma}_{\gamma\gamma}/dtd\varphi$ depends on $\varphi$ due to the Stokes parameters $\xi_1$ and $\xi_3$ [see Figures \ref{fig:dsig} (c)-(e)]. At $\varphi=0$, the distribution of $d\bar{\sigma}_{\gamma\gamma}/dtd\varphi$ along $\cos{\theta_s}$ indicates that reactions are forbidden around the perpendicular direction of the colliding axis. At $\varphi=\pi/4$ and $\varphi=\pi/2$, the distributions of $d\bar{\sigma}_{\gamma\gamma}/dtd\varphi$ are similar to those of $\gamma_R^{(1)}\gamma_R^{(2)}$ and $\gamma_R^{(1)}\gamma_L^{(2)}$ collisions, respectively, implying the superposition state of linear polarization. The total cross section $\sigma_{\gamma\gamma}$ can be decomposed into the LP one with parallel ($\xi_3^{(1)}\xi_3^{(2)}=1$) and mutual-perpendicular ($\xi_3^{(1)}\xi_3^{(2)}=-1$) polarization vectors, and CP one with the same ($\xi_2^{(1)}\xi_2^{(2)}=1$) and opposite ($\xi_2^{(1)}\xi_2^{(2)}=-1$) helicities, as shown in Figure \ref{fig:dsig} (f). The former converges to the nonpolarized case at higher energy $\varepsilon$, while the latter crosses the nonpolarized case at around $\varepsilon=2$ MeV.

The polarization characteristics of LBW $e^+e^-$ pairs are determined by the distributed helicities originating from the linear superpositions of four helicity channels with corresponding weights of $d\sigma_{+-\pm\mp}/d\bar{\sigma}_{\gamma\gamma}$ and $d\sigma_{+-\pm\pm}/d\bar{\sigma}_{\gamma\gamma}$, for example, in $\gamma_{R}^{(1)}\gamma_{L}^{(2)}$ collision. The asymmetric distributions of $d\sigma_{+-+-}/dtd\varphi$ and $d\sigma_{+--+}/dtd\varphi$ indicate the helicity flip along $\cos{\theta_s}$, with $\lambda_+$ varying from -0.5 to 0.5 for $0<\theta_s<\pi$ (a reverse variation for $\lambda_-$), as shown in Fig. \ref{fig:helicity}(a). Consequently, $|+-+-\rangle$ and $|+--+\rangle$ channels contribute to the longitudinal polarization of $e^+e^-$ pairs, as shown in Fig. \ref{fig:helicity}(c). Note that here d$\sigma_{+--+}/dtd\varphi$ is symmetric about $\cos{\theta_s}=0$ with $d\sigma_{+-+-}/dtd\varphi$ and thus is not shown.
On the other hand, the distributions of $d\sigma_{+-++}/dt$ and $d\sigma_{+---}/dt$ are completely overlapping and symmetric along $\cos{\theta_s}$, as shown in Fig. \ref{fig:helicity}(b). Therefore, $|+-++\rangle$ and $|+---\rangle$ channels lead to the equal-weight linear superposition between right-hand and left-hand helicity states and contribute to the purely transverse polarization, as shown in Fig. \ref{fig:helicity}(d). Note that the spin-correlated term in Eq. (\ref{difcs}) has a non-negligible contribution to the $|+-\pm\pm\rangle$ channels and thus to the transverse polarization. In $\gamma_{R}^{(1)}\gamma_{L}^{(2)}$ collision, the distributed $\zeta_{\parallel}$ only originates from $|+-+-\rangle$ and $|+--+\rangle$ channels, while the distributed $\zeta_{\perp}$ originates from all four channels, where $|+-\pm\pm\rangle$ channels contribute to the purely transverse polarization, and the partial overlap between $|+-+-\rangle$ and $|+--+\rangle$ channels leads to the non-equal weights at the helicity states $|\pm\rangle$ and thus the partially transverse polarization.

\begin{figure}[!t]
	\setlength{\abovecaptionskip}{0.2cm}
	\centering\includegraphics[width=1\linewidth]{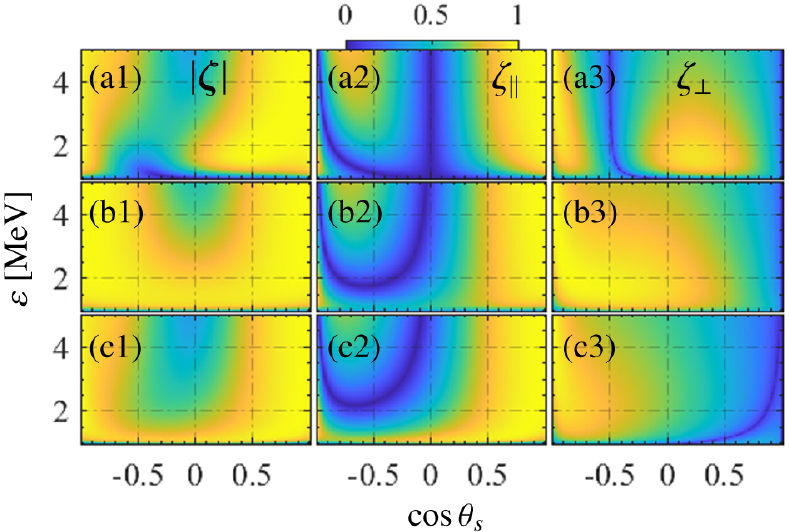}
\caption{Distributions of the total polarization $|\bm\zeta|$, longitudinal polarization $\zeta_\parallel$, and transverse polarization $\zeta_\perp$, calculated from Eq. (\ref{poldef}) with $\xi_{2}^{(1)}=1$ and $\xi_{3}^{(2)}=1$ for different azimuthal angle $\varphi$:  (a1)-(a3) $\varphi=0$, (b1)-(b3) $\varphi=\pi/4$ and (c1)-(c3) $\varphi=\pi/2$. }
	\label{fig:polarization}
\end{figure}
In the colliding scenario of a CP photon and a LP photon, the polarization vector $\bm{\zeta}^{(f)}$ in Eq. (\ref{zeta}) is non-zero due to the Stokes parameter $\xi_2^{(1)}$ and $\varphi$-dependent due to the Stokes parameter $\xi_3^{(2)}$ [see the expression of $G_i^\pm$ in Eqs. (\ref{gele}) and (\ref{gpos})]. Fig. \ref{fig:polarization} shows the distributed polarization at different $\varphi$ angles. As $\varphi$ is rotated from $0$ to $\pi/2$, the distribution of $|\bm{\zeta}|$ wriggles around $\cos{\theta_s}=0$ mainly due to the drastic variation of the distributed $\zeta_\perp$ [see Figs. \ref{fig:polarization}(a3),(b3), and (c3)]. In the distributed $\zeta_\parallel$ and $\zeta_\perp$, the contour line of $\zeta_\parallel=0$ indicates the helicity flip of the pair, while the contour line of $\zeta_\perp=0$ indicates the direction reversal of the transverse spin component $\zeta_{\pm,3}^{(f)}$ due to the vanished $\zeta_{\pm,1}^{(f)}$. In the collision scenario with $\xi_{2}^{(1)}\xi_{1}^{(2)}=1$, there are similar results to those shown in Fig. \ref{fig:polarization}, but with an azimuthal-angle difference of $\Delta\varphi=\pi/4$ due to the fact that $\xi_1$ describes the polarization vector along the direction at an angle of $\pi/4$ in the basis $\bm{e}_1$, and the distributed $\zeta_\perp$ results from both the transverse spin components of $\zeta_{\pm,1}^{(f)}$ and $\zeta_{\pm,3}^{(f)}$.

\subsection{Angle dependence of LBW pairs induced by linear polarization of photon from numerical simulation}
\begin{figure}[!b]
	\setlength{\abovecaptionskip}{0.2cm}
	\centering\includegraphics[width=1\linewidth]{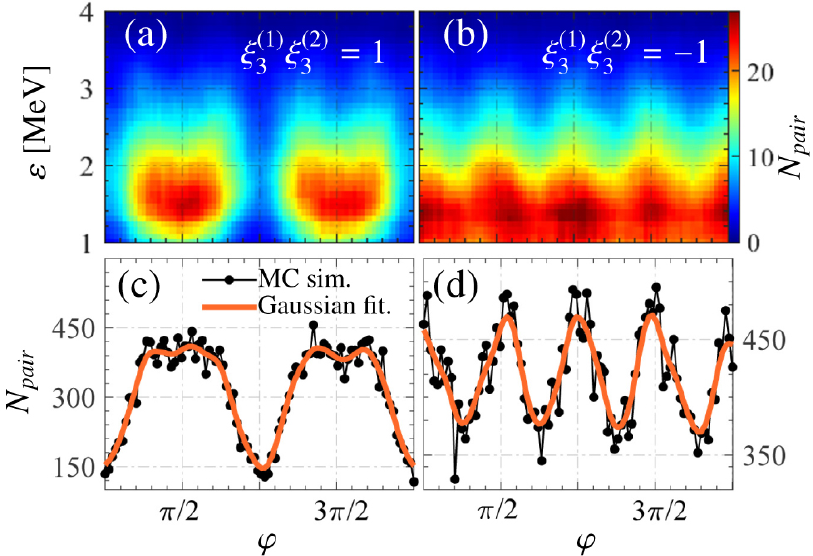}
\caption{ (a) and (b) Distributions of produced number density $N_{pair}$ of LBW pairs for collision scenarios with relatively parallel ($\xi_3^{(1)}\xi_3^{(2)}=1$) and perpendicular ($\xi_3^{(1)}\xi_3^{(2)}=-1$) directions of transverse polarization vectors of photons, respectively. (c) and (d) Statistical variations of produced pair number with azimuthal angle $\varphi$ for two collision scenarios corresponding to (a) and (b), respectively. }
	\label{fig:dsigmc}
\end{figure}
To retrieve the azimuthal and polar dependence of $e^+e^-$ pairs independently, we perform a MC numerical simulation. We initialize approximately $10^{11}$ photons into a bi-gaussian beam with uniform energy distribution between 0.1 MeV-2 MeV. The $\varepsilon-\theta_s$ correlated spectra of LBW pairs is induced only by circular polarization, as previously shown in Ref. \cite{Zhao2022}. Here, we focus on the clarification of $\varphi$-dependence. As indicated in Eqs. (\ref{eta1}) and (\ref{eta2}), the azimuthal dependence of LBW cross section is embodied in the linear polarization of the photon through $\xi_1$ and $\xi_3$. The analytical calculation of $d\bar{\sigma}_{\gamma\gamma}$ with $\xi_3^{(1)}\xi_3^{(2)}=1$ indicates that the produced $e^+e^-$ pairs have an explicit $\varphi$-dependence in the collision of LP photons [see Figs. \ref{fig:dsig}(c)-(e)]. To visualize the azimuthal distribution of the produced $e^+e^-$ pairs, we implement a numerical simulation for the collision of two $\gamma$-photon beams with complete linear polarization. For the collision scenario of $\xi_3^{(1)}\xi_3^{(2)}=1$, the variation of produced pairs along the azimuthal angle presents an approximate $\cos{2\varphi}$-dependence (with two flat-top peaks around $\varphi=\pi/2$ and $\varphi=3\pi/2$) [see Figs. \ref{fig:dsigmc}(a) and (c)]. On the other hand, for the collision scenario of $\xi_3^{(1)}\xi_3^{(2)}=-1$, the variation of produced pairs along the azimuthal angle presents a $\cos{4\varphi}$-dependence [see Figs. \ref{fig:dsigmc}(b) and (d)], which has also been experimentally verified in the collision of high-energy ions \cite{Adam2021}. The distinct $\varphi$-dependence of the produced pairs for the two collision scenarios can be explained by the $\varphi$-dependent cross section [see Eq. (\ref{psidepen})]. Specifically, when $\xi_3^{(1)}=\xi_3^{(2)}=1$, the $\cos{(2\varphi)}$-associated terms are non-vanishing, which dominates the $\varphi$-dependent variation. However, when $\xi_3^{(1)}=1$ and $\xi_3^{(2)}=-1$, the $\cos{(2\varphi)}$-associated terms are vanished, which leads to the only $\cos{(4\varphi)}$-dependent variation. The explicit mathematical expression for this cross section is given as
\begin{align} \label{psidepen}
\frac{d\bar{\sigma}_{\gamma\gamma}}{d\varphi}\propto&\sqrt{(s-4)s}\left[(s+2)\xi _3^{(1)}\xi _3^{(2)}\cos{(4\varphi)}-4(\xi _3^{(1)}+\xi _3^{(2)})\cos{(2\varphi)}\right.\nonumber\\
+&\left.2\xi _3^{(1)}\xi _3^{(2)}\right]+2\tanh^{-1}{\sqrt{(s-4)/s}}\left[-4(s-1)\xi _3^{(1)}\xi _3^{(2)}\cos{(4\varphi)}\right.\nonumber\\
+&\left.4(s-2)(\xi _3^{(1)}+\xi _3^{(2)})\cos{(2\varphi)}+4\xi _3^{(1)}\xi _3^{(2)}\right].
\end{align}

\begin{figure}[!t]
	\setlength{\abovecaptionskip}{0.2cm}
	\centering\includegraphics[width=1\linewidth]{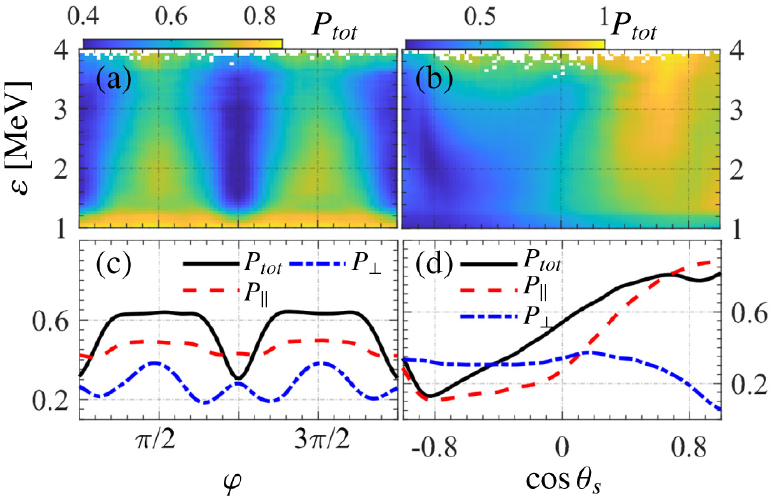}
\caption{ MC simulation for collision scenario with a LP photon beam and a CP photon beam: (a) and (b) Distributions of statistical polarization $P_{tot}$ in $\varepsilon-\varphi$ plane and $\varepsilon-\theta_s$ plane, respectively. (c) and (d) Statistical variations of total polarization $P_{tot}$ of LBW pairs, with longitudinal component $P_\parallel$ and transverse component $P_\perp$, with azimuthal angle $\varphi$ and $\cos{\theta_s}$, respectively. }
	\label{fig:pol-CL}
\end{figure}
In principle, the linear polarization state of a photon is the superposition state of right-hand and left-hand circular polarization states, which implies that polarized $e^+e^-$ pairs can be produced through the collision of a LP photon and a CP photon. Considering the collision scenario with $\xi_2^{(1)}=1$ and $\xi_3^{(2)}=1$, the total polarization $P_{tot}$ of the produced $e^+e^-$ pairs presents both $\varphi$-dependence and $\theta_s$-dependence, as shown in Figs. \ref{fig:pol-CL}(a) and (b). $P_\parallel$ is slightly dependent on $\varphi$, and $P_\perp$ presents prominent $\cos{(2\varphi)}$-dependence, which leads to the flat-top variation of $P_{tot}$ [see Fig. \ref{fig:pol-CL}(c)]. The azimuthal dependence of pair polarization is attributed to the analytical polarization in Eq. (\ref{poldef}) with $\xi_2^{(1)}=1$ and $\xi_3^{(2)}=1$. By substituting Eqs. (\ref{eta1}) and (\ref{eta2}) into the concrete expressions of Eq. (\ref{zeta}), one can find that the spin vectors are $\cos{(2\varphi)}$-dependent in this collision scenario. From the analytical results in Fig. \ref{fig:polarization}, the various polarization distribution at each azimuthal angle implies that when averaging over $\varphi=0-2\pi$ for the produced pairs, the resulting polarization distribution is asymmetric in the $\varepsilon-\theta_s$ plane. The monotonically increasing $P_{\parallel}$ with $\cos{\theta_s}$ together with the slow-varying $P_{\perp}$ leads to the almost linearly increasing $P_{tot}$, as shown in Fig. \ref{fig:pol-CL}(d).

\section{Discussion}\label{discussion}
Non-polarized $\gamma$-photon beams with ultrahigh peak brilliance exceeding $10^{20}$ $\rm{photons~s^{-1}mm^{-2}mrad^{-2} 0.1\% BW}$ can be generated experimentally via nonlinear Thomson scattering with monotonically decreasing spectra \cite{Sarri2014,Yan2017} and inverse Compton scattering quasimonochromatic spectra \cite{Albert2010,Yu2016}. Although the required brilliance for producing LBW pairs into $\mu$A-order current in a single shot is approached by such high brilliance \cite{Zhao2022}, the interaction of $\gamma$-photons with experimental equipment results in significant background noise signals through more efficient Bethe-Heitler and triplet pair productions \cite{Ribeyre2016,Yu2019}, which consequently reduces the signal-to-noise ratio. In principle, the signal-to-noise ratio in detecting the BW process can be enhanced by designing high-density $\gamma$-photon sources while ensuring the interactions in a vacuum environment. However, the production of background noise is inevitable and increases with the density of $\gamma$-photons. Hence, verifying the BW process through increasing intensity signature remains challenging in $\gamma\gamma$ colliders.

Recently, several proposals have been put forward to generate LP or CP $\gamma$-photon beams with brilliances of around $10^{21}$ $\rm{photons~s^{-1}mm^{-2}mrad^{-2} 0.1\% BW}$ through nonlinear Compton scattering \cite{Xue2020,Li2020polarized,Tang2020,Wang2022}. Laser-irradiated ultrathin planar aluminum targets can produce $\gamma$ photons with an average linear polarization of $70\%$ \cite{Xue2020}. Electrons colliding with lasers can generate multi-GeV CP (LP) $\gamma$ photons with a polarization of about $90\%$ \cite{Li2020polarized,Tang2020}. Additionally, a weakly nonlinear Compton scattering scheme that combines laser plasma wakefield acceleration and plasma mirror techniques can produce CP $\gamma$-photon beams with a polarization of about $61\%$ \cite{Wang2022}. Thus, there is potential to use a polarized $\gamma\gamma$ collider to verify the LBW process by detecting the spin polarization of $e^\pm$ produced by collisions of CP photons \cite{Abbott2016,Grames2020} or by detecting the azimuthal angle-dependent pair yield produced by collisions of LP photons \cite{Adam2021,Kettle2021}. Hence, the qualitatively polarization-associated signatures are likely to be more distinguishable when compared to the single intensity signature of $e^\pm$ pairs. Overall, the use of ultrabrilliant and polarized $\gamma$-photon sources presents new prospects for detecting the polarization of LBW pairs.


\section{CONCLUSION}\label{summary}
In summary, we investigate the fully angle-resolved LBW process using the completely polarized cross-section and MC numerical simulation. The analytical longitudinal and transverse polarization are decoupled by the newly-defined spin base of $e^\pm$, which allows for the analytical clarification of the pair polarization mechanism through the differential cross-section with different helicity channels.

The fully spin-resolved MC method is utilized to simulate the polarized $e^+e^-$ pair productions in the realistic $\gamma$-photon beam-beam collision, wherein the $\cos{2\varphi}$ and $\cos{4\varphi}$-modulated pair yield is verified for the relatively parallel and perpendicular polarization vectors of two colliding LP photons, respectively. When colliding with the CP photons, the linear polarization of photon also induces the $\cos{2\varphi}$-modulated and asymmetrically $\cos{\theta_s}$-dependent polarization of $e^+e^-$ pairs.

The environment of real photon-photon collisions provides an opportunity to test basic physics and elucidate associated high-energy astrophysics. Therefore, further research is needed to explore the polarization effects in photon-photon collisions and their implications for our understanding of fundamental physics and astrophysical processes.

\section{ACKNOWLEDGEMENT}
The work is supported by the National Natural Science Foundation of China (Grants Nos. 11874295, 12022506, U2267204, 12105217), the Open Fund of the State Key Laboratory of High Field Laser Physics (Shanghai Institute of Optics and Fine Mechanics), and the foundation of science and technology on plasma physics laboratory (no. JCKYS2021212008).

\appendix
\section{The expressions of the total cross section and  spin-summarized differential cross section } \label{appA}
The coefficients $f_1,f_2,f_3$ and $f_4$ in Eq. (\ref{ebase}) are
	\begin{subequations}
		\begin{align}
		f_1&=\frac{-x-y+2}{\sqrt{(x+y-4) (x+y)}}, \\
		f_2&=\frac{2}{\sqrt{(x+y-4) (x+y)}}, \\
		f_3&=-\frac{2 \sqrt{x y-x-y}}{\sqrt{x+y-4} (x+y)}, \\
		f_4&=\frac{y-x}{2 \sqrt{x+y-4} \sqrt{x y-x-y}}.
	\end{align}
\end{subequations}
with the normalized variants $x=1-t$ and $y=1-u$.

The expressions of the spin-summarized differential cross section is
\begin{eqnarray}\label{dsig_spinsum}
\frac{d^2\bar{\sigma}_{\gamma\gamma}}{dtd\varphi}=\frac{r_e^2m_e^2}{64\varepsilon^4}(F_0+F_1)
\end{eqnarray}
where
\begin{widetext}
\begin{subequations}
	\begin{align}
		F_0&=\frac{4 \left(x^3 y+4 x^2 y-4 x^2+x y^3+4 x y^2-8 x y-4 y^2\right)}{x^2 y^2}, \\
		F_1&=\frac{4}{x^2 y^2}\times\nonumber\\
		&\left[2(x(-y)+x+y)^2cos(4\varphi)(\xi_1^{(1)}\xi_1^{(2)}-\xi_3^{(1)}\xi_3^{(2)}))-2 \xi _1^{(1)} \xi _1^{(2)} x^2-4 \xi _1^{(1)} \xi _1^{(2)} x y-2 \xi _1^{(1)} \xi _1^{(2)} y^2-2 \xi _1^{(1)} \xi _3^{(2)} x^2 sin(4 \varphi)-2 \xi _1^{(1)} \xi _3^{(2)} x^2 y^2 sin(4 \varphi )\right.\nonumber\\
		&+4 \xi _1^{(1)} \xi _3^{(2)} x^2 y sin(4 \varphi )+4 \xi _1^{(1)} \xi _3^{(2)} x y^2 sin(4 \varphi )-4 \xi _1^{(1)} \xi _3^{(2)} x y sin(4 \varphi )-2 \xi _1^{(1)} \xi _3^{(2)} y^2 sin(4 \varphi )+4 \xi _1^{(1)} x^2 sin(2 \varphi )-4 \xi _1^{(1)} x^2 y sin(2 \varphi )\nonumber\\
		&-4 \xi _1^{(1)} x y^2 sin(2 \varphi )+8 \xi _1^{(1)} x y sin(2 \varphi )+4 \xi _1^{(1)} y^2 sin(2 \varphi )-2 \xi _1^{(2)} \xi _3^{(1)} x^2 sin(4 \varphi )-2 \xi _1^{(2)} \xi _3^{(1)} x^2 y^2 sin(4 \varphi )+4 \xi _1^{(2)} \xi _3^{(1)} x^2 y sin(4 \varphi )\nonumber\\
		&+4 \xi _1^{(2)} \xi _3^{(1)} x y^2 sin(4 \varphi )-4 \xi _1^{(2)} \xi _3^{(1)} x y sin(4 \varphi )-2 \xi _1^{(2)} \xi _3^{(1)} y^2 sin(4 \varphi )+4 \xi _1^{(2)} x^2 sin(2 \varphi )-4 \xi _1^{(2)} x^2 y sin(2 \varphi )-4 \xi _1^{(2)} x y^2 sin(2 \varphi )+8 \xi _1^{(2)} x y sin(2 \varphi )\nonumber\\
		&+4 \xi _1^{(2)} y^2 sin(2 \varphi )+\xi _2^{(1)} \xi _2^{(2)} x^3 y-2 \xi _2^{(1)} \xi _2^{(2)} x^3-2 \xi _2^{(1)} \xi _2^{(2)} x^2 y+\xi _2^{(1)} \xi _2^{(2)} x y^3-2 \xi _2^{(1)} \xi _2^{(2)} x y^2-2 \xi _2^{(1)} \xi _2^{(2)} y^3\nonumber\\
		&\left.-4 \left(\xi _3^{(2)}+\xi _3^{(1)}\right) \left(x^2 (y-1)+x (y-2) y-y^2\right) cos(2 \varphi )-2 \xi _3^{(1)} \xi _3^{(2)} x^2-4 \xi _3^{(1)} \xi _3^{(2)} x y-2 \xi _3^{(1)} \xi _3^{(2)} y^2\right].
	\end{align}
\end{subequations}
\end{widetext}

The expression of total cross section $\sigma_{\gamma\gamma}$ is
\begin{align}\label{sig_tot}
\sigma_{\gamma\gamma}=\frac{r_e^2m_e^4}{64\varepsilon^4}\tilde{F},
\end{align}
where $\tilde{F}$ is the integration of $F$ with $s=4\varepsilon^2/m_e^2$:
\begin{align}
\tilde{F}=&16\pi\sqrt{\frac{s-4}{s}}\left( -s -4+2\xi _1^{(1)}\xi _1^{(2)}+3s \xi _2^{(1)}\xi _2^{(2)}-2 \xi _3^{(1)}\xi _3^{(2)}\right)\nonumber\\
&+\frac{16\pi}{s}\tanh^{-1}{\sqrt{\frac{s-4}{s}}}\left(2s^2+8s-16\right.\nonumber\\
&\left.+8 \xi _1^{(1)}\xi _1^{(2)}-2 s^2\xi _2^{(1)}\xi _2^{(2)}-8\xi _3^{(1)}\xi _3^{(2)}\right).
\end{align}

\section{The expressions of the coefficients in Eq. (\ref{difcs})} \label{appB}

The final-state-spin irrelevant term of Eq. (\ref{difcs}) is
\begin{widetext}
\begin{eqnarray}
F=&-\frac{8 (x (-y)+2 x+2 y)}{x y}\xi^{(1)}_1\xi^{(2)}_1-\frac{4(x(-y)+2x+2y)\left(x^2+y^2\right)}{x^2 y^2}\xi^{(1)}_2\xi^{(2)}_2-\frac{8\left(x^2y^2-2x^2y+2x^2-2xy^2+4xy+2y^2\right)}{x^2y^2}\xi^{(1)}_3\xi^{(2)}_3\nonumber\\
&+\frac{16(x+y)(x(-y)+x+y)}{x^2y^2}(\xi^{(1)}_3+\xi^{(2)}_3)
\end{eqnarray}
\end{widetext}

The coefficients of electron spin $\zeta_{-,i}$ in Eq. (\ref{difcs}) are
\begin{widetext}
\begin{subequations}\label{gele}
	\begin{align}
	G^-_{1} &=-\frac{8(x+y)\sqrt{x^2y-x^2+xy^2-2xy-y^2}}{x^2y^2}\left(\xi_1^{(1)}\xi_2^{(2)}x-\xi_1^{(2)}\xi_2^{(1)}y\right),\\
	G^-_{2} &=\frac{4 (x+y) \sqrt{x^2+2 x y-4 x+y^2-4 y}}{x^2 y^2 (x+y-4)}\times\nonumber\\
	&\left(\xi_2^{(1)} x^2 y+4 \xi _2^{(1)} \xi _3^{(2)} x-4 \xi _2^{(1)} \xi _3^{(2)} x y+4 \xi _2^{(1)} \xi _3^{(2)}y-4 \xi _2^{(1)} x-4 \xi _2^{(2)} \xi _3^{(1)} x-\xi _2^{(1)} x y^2+2 \xi _2^{(1)} x y+4 \xi _2^{(2)} \xi _3^{(1)} x y+2 \xi _2^{(1)} y^2-4 \xi _2^{(1)} y-2 \xi _2^{(2)} x^2\right.\nonumber\\
	&\left.+\xi _2^{(2)} x^2 y+4 \xi _2^{(2)} x-\xi _2^{(2)} x y^2-2 \xi _2^{(2)} x y+4 \xi _2^{(2)} y-4 \xi _3^{(1)} \xi _2^{(2)} y\right) ,\\
	G^-_{3} &=-\frac{1}{x^2 y^2 \sqrt{x+y-4} \sqrt{x y-x-y}}\times\nonumber\\
	&8\left[\xi _2^{(1)} \xi _3^{(2)} x^3 y^2-3 \xi _2^{(1)} \xi _3^{(2)} x^3 y+2 \xi _2^{(1)} \xi _3^{(2)} x^3+2 \xi _2^{(1)} \xi _3^{(2)} x^2 y^3-7 \xi _2^{(1)} \xi _3^{(2)} x^2 y^2+6 \xi _2^{(1)} \xi _3^{(2)} x^2 y+\xi _2^{(1)} \xi _3^{(2)} x y^4-5 \xi _2^{(1)} \xi _3^{(2)} x y^3+6 \xi _2^{(1)} \xi _3^{(2)} x y^2\right.\nonumber\\
	&-\xi _2^{(1)} \xi _3^{(2)} y^4+\xi _2^{(2)} \left(-\xi _3^{(1)}\right) x^4+\xi _2^{(2)} \xi _3^{(1)} x^4 y-2 \xi _2^{(1)} x^3-2 \xi _2^{(1)} x^3 y^2+4 \xi _2^{(1)} x^3 y-2 \xi _2^{(1)} x^2 y^3+8 \xi _2^{(1)} x^2 y^2-6 \xi _2^{(1)} x^2 y+4 \xi _2^{(1)} x y^3-6 \xi _2^{(1)} x y^2\nonumber\\
	&-2 \xi _2^{(1)} y^3+2 \xi _2^{(1)} \xi _3^{(2)} y^3+2 \xi _2^{(2)} \xi _3^{(1)} x^3-2 \xi _2^{(2)} x^3 y^2+2 \xi _2^{(2)} \xi _3^{(1)} x^3 y^2+4 \xi _2^{(2)} x^3 y-5 \xi _2^{(2)} \xi _3^{(1)} x^3 y+\xi _2^{(2)} \xi _3^{(1)} x^2 y^3-7 \xi _2^{(2)} \xi _3^{(1)} x^2 y^2+6 \xi _2^{(2)} \xi _3^{(1)} x^2 y\nonumber\\
	&\left.-3 \xi _2^{(2)} \xi _3^{(1)} x y^3+6 \xi _2^{(2)} \xi _3^{(1)} x y^2+2 \xi _2^{(2)} \xi _3^{(1)} y^3-2 \xi _2^{(2)} x^3-2 \xi _2^{(2)} x^2 y^3+8 \xi _2^{(2)} x^2 y^2-6 \xi _2^{(2)} x^2 y+4 \xi _2^{(2)} x y^3-6 \xi _2^{(2)} x y^2-2 \xi _2^{(2)} y^3\right]
	\end{align}
\end{subequations}
\end{widetext}
The coefficients of positron spin $\zeta_{+,i}$ in Eq. (\ref{difcs}) are
\begin{widetext}
\begin{subequations}\label{gpos}
	\begin{align}		
	G^+_{1} &=-\frac{8 (x+y) \sqrt{-(x+y) (x (-y)+x+y)} }{x^2 y^2}\left(\xi _1^{(2)} \xi _2^{(1)} x-\xi _1^{(1)} \xi _2^{(2)} y\right), \\
	G^+_{2} &=-\frac{4 (x+y) \sqrt{x^2+2 x y-4 x+y^2-4 y}}{x^2 y^2 (x+y-4)}\times\nonumber\\
	&\left(-2 \xi _2^{(1)} x^2+\xi _2^{(1)} x^2 y-4 \xi _2^{(1)} \xi _3^{(2)} x+4 \xi _2^{(1)} x+4 \xi _2^{(2)} \xi _3^{(1)} x-\xi _2^{(1)} x y^2+4 \xi _2^{(1)} \xi _3^{(2)} x y-2 \xi _2^{(1)} x y-4 \xi _2^{(2)} \xi _3^{(1)} x y-4 \xi _2^{(1)} \xi _3^{(2)} y+4 \xi _2^{(1)} y+4 \xi _2^{(2)} \xi _3^{(1)} y\right.\nonumber\\
	&\left.+\xi _2^{(2)} x^2 y-4 \xi _2^{(2)} x-\xi _2^{(2)} x y^2+2 \xi _2^{(2)} x y+2 \xi _2^{(2)} y^2-4 \xi _2^{(2)} y\right),\\
	G^+_{3} &=-\frac{1}{x^2 y^2 \sqrt{x+y-4}}8(x+y)\sqrt{x y-x-y}\times\nonumber\\
	&\left(\xi _3^{(2)} \xi _2^{(1)} x^2-2 \xi _3^{(1)} \xi _2^{(2)} x+2 \xi _2^{(2)} x-2 \xi _3^{(2)} \xi _2^{(1)} x+2 \xi _2^{(1)} x-2 \xi _2^{(2)} x y+\xi _3^{(1)} \xi _2^{(2)} x y+\xi _3^{(2)} \xi _2^{(1)} x y-2 \xi _2^{(1)} x y+\xi _3^{(1)} \xi _2^{(2)} y^2-2 \xi _3^{(1)} \xi _2^{(2)} y+2 \xi _2^{(2)} y\right.\nonumber\\
	&\left.-2 \xi _3^{(2)} \xi _2^{(1)} y+2 \xi _2^{(1)} y\right).
    \end{align}
\end{subequations}
\end{widetext}

The coefficients of spin-correlated term $\zeta_{-,i}\zeta_{+,j}$ in Eq. (\ref{difcs}) are
\begin{widetext}
\begin{subequations}
	\begin{align}
	H_{11}&=\frac{1}{x^2 y^2}4\left(-\xi _1^{(1)} \xi _1^{(2)} x^3 y+2 \xi _1^{(1)} \xi _1^{(2)} x^3-2 \xi _1^{(1)} \xi _1^{(2)} x^2 y^2+3 \xi _1^{(1)} \xi _1^{(2)} x^2 y-\xi _1^{(1)} \xi _1^{(2)} x y^3+2 \xi _1^{(1)} \xi _1^{(2)} x y^2+2 \xi _1^{(1)} \xi _1^{(2)} y^3+3 \xi _2^{(1)} \xi _2^{(2)} x^2 y\right.\nonumber\\
	&+4 \xi _2^{(1)} \xi _2^{(2)} x y^2+\xi _3^{(1)} \xi _3^{(2)} x^3 y-2 \xi _3^{(1)} \xi _3^{(2)} x^2 y^2+5 \xi _3^{(1)} \xi _3^{(2)} x^2 y-4 \xi _3^{(1)} \xi _3^{(2)} x^2+\xi _3^{(1)} \xi _3^{(2)} x y^3+4 \xi _3^{(1)} \xi _3^{(2)} x y^2-8 \xi _3^{(1)} \xi _3^{(2)} x y-4 \xi _3^{(1)} \xi _3^{(2)} y^2\nonumber\\
	&+4 \xi _3^{(2)} x^2+4 \xi _3^{(1)} x^2-4 \xi _3^{(2)} x^2 y-4 \xi _3^{(1)} x^2 y-4 \xi _3^{(2)} x y^2-4 \xi _3^{(1)} x y^2+8 \xi _3^{(2)} x y+8 \xi _3^{(1)} x y+4 \xi _3^{(2)} y^2+4 \xi _3^{(1)} y^2\nonumber\\
	&\left.-2 x^2 y^2+4 x^2 y-4 x^2+4 x y^2-8 x y-4 y^2\right) , \\		
	H_{12}&=-\frac{4  (x+y) \sqrt{-(x+y) (x (-y)+x+y)} \sqrt{x^2+2 x y-4 x+y^2-4 y}}{x^2 y^2}\left(\xi _1^{(2)}+\xi _1^{(1)}\right) , \\
	H_{13}&=\frac{4  \sqrt{x+y-4} (x+y)^2 \sqrt{-(x+y) (x (-y)+x+y)}}{x^2 y^2 \sqrt{x y-x-y}}\left(\xi _1^{(2)} \xi _3^{(1)}-\xi _1^{(1)} \xi _3^{(2)}\right) ,
\end{align}
\end{subequations}
\end{widetext}

\begin{widetext}
\begin{subequations}
\begin{align}
	H_{21}&=\frac{2  (x+y) \sqrt{-(x+y) (x (-y)+x+y)} \sqrt{x^2+2 x y-4 x+y^2-4 y}}{x^2 y^2}\left(\xi _1^{(2)}+\xi _1^{(1)}\right) ,\\
	H_{22}&=-\frac{1}{x^2 y^2 (x+y-4)}\times\nonumber\\
	&4\left(\xi _2^{(1)} \xi _2^{(2)} x^4 y+4 \xi _1^{(1)} \xi _1^{(2)} x^3-4 \xi _2^{(1)} \xi _2^{(2)} x^3-4 \xi _3^{(2)} x^3-4 \xi _3^{(1)} x^3-2 \xi _3^{(1)} \xi _3^{(2)} x^3 y^2+2 \xi _1^{(1)} \xi _1^{(2)} x^3 y^2+\xi _2^{(1)} \xi _2^{(2)} x^3 y^2+4 \xi _3^{(2)} x^3 y+4 \xi _3^{(1)} \xi _3^{(2)} x^3 y\right.\nonumber\\
	&-4 \xi _1^{(1)} \xi _1^{(2)} x^3 y+4 \xi _3^{(1)} x^3 y-16 \xi _3^{(1)} \xi _3^{(2)} x^2+16 \xi _3^{(2)} x^2+16 \xi _3^{(1)} x^2-2 \xi _3^{(1)} \xi _3^{(2)} x^2 y^3+2 \xi _1^{(1)} \xi _1^{(2)} x^2 y^3+\xi _2^{(1)} \xi _2^{(2)} x^2 y^3-8 \xi _2^{(1)} \xi _2^{(2)} x^2 y^2+8 \xi _3^{(2)} x^2 y^2\nonumber\\
	&+8 \xi _3^{(1)} x^2 y^2+4 \xi _2^{(1)} \xi _2^{(2)} x^2 y+16 \xi _3^{(1)} \xi _3^{(2)} x^2 y-4 \xi _1^{(1)} \xi _1^{(2)} x^2 y-28 \xi _3^{(2)} x^2 y-28 \xi _3^{(1)} x^2 y+\xi _2^{(1)} \xi _2^{(2)} x y^4-4 \xi _1^{(1)} \xi _1^{(2)} x y^3+4 \xi _3^{(2)} x y^3+4 \xi _3^{(1)} \xi _3^{(2)} x y^3\nonumber\\
	&+4 \xi _3^{(1)} x y^3-4 \xi _1^{(1)} \xi _1^{(2)} x y^2-28 \xi _3^{(2)} x y^2-28 \xi _3^{(1)} x y^2+4 \xi _2^{(1)} \xi _2^{(2)} x y^2+16 \xi _3^{(1)} \xi _3^{(2)} x y^2+32 \xi _3^{(2)} x y-32 \xi _3^{(1)} \xi _3^{(2)} x y+32 \xi _3^{(1)} x y-4 \xi _2^{(1)} \xi _2^{(2)} y^3\nonumber\\
	&-4 \xi _3^{(2)} y^3-4 \xi _3^{(1)} y^3+4 \xi _1^{(1)} \xi _1^{(2)} y^3-16 \xi _3^{(1)} \xi _3^{(2)} y^2+16 \xi _3^{(2)} y^2+16 \xi _3^{(1)} y^2+x^4 y-2 x^4+x^3 y^2-8 x^3 y+8 x^3+x^2 y^3-20 x^2 y^2+40 x^2 y\nonumber\\
	&\left.-16 x^2+x y^4-8 x y^3+40 x y^2-32 x y-2 y^4+8 y^3-16 y^2\right),\\
H_{23}&=\frac{4 \sqrt{x y-x-y} \sqrt{x^2+2 x y-4 x+y^2-4 y}}{x^2 y^2 (x+y-4)^{3/2}}\nonumber\\
	&\left(\xi _3^{(2)} x^3+\xi _3^{(1)} x^3-4 \xi _1^{(1)} \xi _1^{(2)} x^2+4 \xi _2^{(1)} \xi _2^{(2)} x^2+4 \xi _3^{(1)} \xi _3^{(2)} x^2-8 \xi _3^{(1)} x^2+4 \xi _1^{(1)} \xi _1^{(2)} x^2 y-4 \xi _2^{(1)} \xi _2^{(2)} x^2 y-4 \xi _3^{(1)} \xi _3^{(2)} x^2 y-\xi _3^{(2)} x^2 y+3 \xi _3^{(1)} x^2 y\right.\nonumber\\
	&-4 \xi _1^{(1)} \xi _1^{(2)} x y^2-4 \xi _3^{(2)} x y^2+4 \xi _2^{(1)} \xi _2^{(2)} x y^2+4 \xi _3^{(1)} \xi _3^{(2)} x y^2+2 \xi _3^{(1)} x y^2+8 \xi _3^{(2)} x y-8 \xi _3^{(1)} x y-2 \xi _3^{(2)} y^3-4 \xi _2^{(1)} \xi _2^{(2)} y^2-4 \xi _3^{(1)} \xi _3^{(2)} y^2+4 \xi _1^{(1)} \xi _1^{(2)} y^2\nonumber\\
	&\left.+8 \xi _3^{(2)} y^2-4 x^2 y+4 x^2+4 x y^2-4 y^2\right) ,
\end{align}
\end{subequations}
\end{widetext}

\begin{widetext}
\begin{subequations}
	\begin{align}
	H_{31}&=-\frac{2 (x+y) \sqrt{-(x+y) (x (-y)+x+y)} }{x^2 y^2 \sqrt{x+y-4} \sqrt{x y-x-y}}\times\nonumber\\
	&\left(2 \xi _3^{(1)} \xi _1^{(2)} x^2-2 \xi _3^{(2)} \xi _1^{(1)} x^2-8 \xi _3^{(1)} \xi _1^{(2)} x+8 \xi _3^{(2)} \xi _1^{(1)} x-\xi _3^{(2)} \xi _1^{(1)} x y^2+4 \xi _3^{(1)} \xi _1^{(2)} x y-4 \xi _3^{(2)} \xi _1^{(1)} x y+2 \xi _3^{(1)} \xi _1^{(2)} y^2-\xi _3^{(2)} \xi _1^{(1)} y^2-8 \xi _3^{(1)} \xi _1^{(2)} y\right.\nonumber\\
	&\left.+8 \xi _3^{(2)} \xi _1^{(1)} y\right) ,\\
	H_{32}&=-\frac{4 \sqrt{x y-x-y} \sqrt{x^2+2 x y-4 x+y^2-4 y}}{x^2 y^2 (x+y-4)^{3/2}}\times\nonumber\\
	&\left(\xi _3^{(2)} x^3+\xi _3^{(1)} x^3-4 \xi _1^{(1)} \xi _1^{(2)} x^2+4 \xi _2^{(1)} \xi _2^{(2)} x^2+4 \xi _3^{(1)} \xi _3^{(2)} x^2-8 \xi _3^{(2)} x^2+4 \xi _1^{(1)} \xi _1^{(2)} x^2 y+3 \xi _3^{(2)} x^2 y-4 \xi _2^{(1)} \xi _2^{(2)} x^2 y-4 \xi _3^{(1)} \xi _3^{(2)} x^2 y-\xi _3^{(1)} x^2 y\right.\nonumber\\
	&-4 \xi _1^{(1)} \xi _1^{(2)} x y^2-4 \xi _3^{(1)} x y^2+4 \xi _2^{(1)} \xi _2^{(2)} x y^2+2 \xi _3^{(2)} x y^2+4 \xi _3^{(1)} \xi _3^{(2)} x y^2-8 \xi _3^{(2)} x y+8 \xi _3^{(1)} x y-2 \xi _3^{(1)} y^3-4 \xi _2^{(1)} \xi _2^{(2)} y^2-4 \xi _3^{(1)} \xi _3^{(2)} y^2+4 \xi _1^{(1)} \xi _1^{(2)} y^2\nonumber\\
	&\left.+8 \xi _3^{(1)} y^2-4 x^2 y+4 x^2+4 x y^2-4 y^2\right) ,\\
	H_{33}&=-\frac{1}{x^2 y^2 (x+y-4)}\times\nonumber\\
	&4\left(-2 \xi _3^{(1)} \xi _3^{(2)} x^4+\xi _3^{(1)} \xi _3^{(2)} x^4 y-\xi _1^{(1)} \xi _1^{(2)} x^4 y+4 \xi _1^{(1)} \xi _1^{(2)} x^3-4 \xi _2^{(1)} \xi _2^{(2)} x^3+8 \xi _3^{(1)} \xi _3^{(2)} x^3-4 \xi _3^{(2)} x^3-4 \xi _3^{(1)} x^3-\xi _1^{(1)} \xi _1^{(2)} x^3 y^2-2 \xi _2^{(1)} \xi _2^{(2)} x^3 y^2\right.\nonumber\\
	&+\xi _3^{(1)} \xi _3^{(2)} x^3 y^2+4 \xi _2^{(1)} \xi _2^{(2)} x^3 y+4 \xi _3^{(2)} x^3 y-8 \xi _3^{(1)} \xi _3^{(2)} x^3 y+4 \xi _3^{(1)} x^3 y-16 \xi _3^{(1)} \xi _3^{(2)} x^2+16 \xi _3^{(2)} x^2+16 \xi _3^{(1)} x^2-\xi _1^{(1)} \xi _1^{(2)} x^2 y^3-2 \xi _2^{(1)} \xi _2^{(2)} x^2 y^3\nonumber\\
	&+\xi _3^{(1)} \xi _3^{(2)} x^2 y^3-20 \xi _3^{(1)} \xi _3^{(2)} x^2 y^2+8 \xi _1^{(1)} \xi _1^{(2)} x^2 y^2+8 \xi _3^{(2)} x^2 y^2+8 \xi _3^{(1)} x^2 y^2+4 \xi _2^{(1)} \xi _2^{(2)} x^2 y+40 \xi _3^{(1)} \xi _3^{(2)} x^2 y-4 \xi _1^{(1)} \xi _1^{(2)} x^2 y-28 \xi _3^{(2)} x^2 y\nonumber\\
	&-28 \xi _3^{(1)} x^2 y-\xi _1^{(1)} \xi _1^{(2)} x y^4+\xi _3^{(1)} \xi _3^{(2)} x y^4-8 \xi _3^{(1)} \xi _3^{(2)} x y^3+4 \xi _2^{(1)} \xi _2^{(2)} x y^3+4 \xi _3^{(2)} x y^3+4 \xi _3^{(1)} x y^3-4 \xi _1^{(1)} \xi _1^{(2)} x y^2-28 \xi _3^{(2)} x y^2-28 \xi _3^{(1)} x y^2\nonumber\\
	&+4 \xi _2^{(1)} \xi _2^{(2)} x y^2+40 \xi _3^{(1)} \xi _3^{(2)} x y^2+32 \xi _3^{(2)} x y-32 \xi _3^{(1)} \xi _3^{(2)} x y+32 \xi _3^{(1)} x y-2 \xi _3^{(1)} \xi _3^{(2)} y^4-4 \xi _2^{(1)} \xi _2^{(2)} y^3-4 \xi _3^{(2)} y^3-4 \xi _3^{(1)} y^3+4 \xi _1^{(1)} \xi _1^{(2)} y^3\nonumber\\
	&+8 \xi _3^{(1)} \xi _3^{(2)} y^3-16 \xi _3^{(1)} \xi _3^{(2)} y^2+16 \xi _3^{(2)} y^2+16 \xi _3^{(1)} y^2-2 x^3 y^2+4 x^3 y-2 x^2 y^3+16 x^2 y-16 x^2+4 x y^3+16 x y^2-32 x y\nonumber\\
	&\left.-16 y^2\right)
    \end{align}
\end{subequations}
\end{widetext}

\bibliography{refs}

\end{document}